\documentclass[a4paper,11pt]{article}
\pdfoutput=1

\usepackage{jheppub}

\usepackage{amsmath,latexsym,amssymb,slashed}
\usepackage{graphicx}
\usepackage{psfrag}
\usepackage[latin1]{inputenc}
\usepackage{subfigure}
\usepackage{bbm}
\usepackage{color}

\newcommand\cm{\mathcal{M}}

\newcommand\beal{\begin{align}}
\newcommand\nn{\nonumber}

\newcommand{\eq}[1]{\begin{equation}#1\end{equation}}
\newcommand{\spl}[1]{\begin{split}#1\end{split}}
\newcommand{\al}[1]{\begin{align}#1\end{align}}
\newcommand{\all}[1]{\begin{align*}#1\end{align*}}

\newcommand{\mcal}{\mathcal{M}}

\newcommand{\ncal}{\mathcal{N}}

\newcommand{\G}{\Gamma}
\newcommand{\g}{\gamma}
\newcommand{\e}{\epsilon}
\newcommand{\we}{\widetilde{\eta}}

\newcommand{\p}{\partial}
\renewcommand{\O}{\Omega}

\renewcommand{\a}{\alpha}

\renewcommand{\L}{\Lambda}

\renewcommand{\t}{\theta}

\newcommand{\boxedeq}[1]{
\begin{equation}
\fbox{
\rule[0.7cm]{0pt}{0pt}
$#1$
\rule[-0.45cm]{0pt}{0pt}
}
\end{equation}
}

\def\d{\text{d}}

\def\slashchar#1{\setbox0=\hbox{$#1$}           
\dimen0=\wd0                                 
\setbox1=\hbox{/} \dimen1=\wd1               
\ifdim\dimen0>\dimen1                        
\rlap{\hbox to \dimen0{\hfil/\hfil}}      
#1                                        
\else                                        
\rlap{\hbox to \dimen1{\hfil$#1$\hfil}}   
/                                         
\fi}

\title{Generalized complex geometry of pure backgrounds\\ in ten and eleven dimensions}

\author{Dani\"{e}l Prins and}
\author{Dimitrios Tsimpis}

\affiliation{Universit\'{e} de Lyon\\
UMR 5822, CNRS/IN2P3, Institut de Physique Nucl\'{e}aire de Lyon\\
4 rue Enrico Fermi,
F-69622 Villeurbanne Cedex,  France}
\emailAdd{dlaprins@ipnl.in2p3.fr}
\emailAdd{tsimpis@ipnl.in2p3.fr}

\abstract{
Pure backgrounds are a natural generalization of supersymmetric Calabi-Yau compactifications in the presence of flux. They are  described in
the language of generalized $SU(d)\times SU(d)$ structures and generalized complex geometry, and they exhibit some interesting general patterns: the internal manifold is generalized Calabi-Yau, while the Ramond-Ramond flux is exact in a precise sense discussed in this paper. We have shown that although these two characteristics do persist in the case of generic ten-dimensional Euclidean type II pure backgrounds, they do not capture the full content of supersymmetry.
We also discuss the uplift of real Euclidean type IIA pure backgrounds to supersymmetric backgrounds of Lorentzian eleven-dimensional supergravity.
}

\begin{document}
\maketitle
\flushbottom
\setcounter{footnote}{0}
\renewcommand{\thefootnote}{\arabic{footnote}}
\setcounter{section}{0}

\section{Introduction and Summary}\label{sec:introduction}

The idea that there is a certain correspondence between supersymmetric solutions and supersymmetric sources has existed for some time \cite{berg1}. With the advent of generalized geometry \cite{hitch,gual} (see \cite{koer} for a review), this statement was made precise and proven in general for type II supergravity backgrounds of the form $\mathbb{R}^{1,3}\times\cm_6$. Specifically, the conditions for an $\ncal=1$ supersymmetric bosonic background of this form can be expressed as a set of first-order differential equations for two complex pure spinors of $Cl(6,6)$ \cite{gran}. In \cite{calmart} it was then shown that these pure-spinor equations are nothing but the differential conditions obeyed by the (generalized) calibration forms of all admissible supersymmetric static, magnetic D-branes in that background.

In \cite{patalong} the one-to-one correspondence between supersymmetry and D-brane calibrations was shown to also hold for type II $\ncal=1$ backgrounds of the form $\mathbb{R}^{1,5}\times\cm_4$. In that reference it was also conjectured that the correpondence should extend to supersymmetric backgrounds of the form $\mathbb{R}^{1,1}\times\cm_8$ for the case where the internal parts (along $\cm_8$) of the Killing spinors are pure spinors of $Cl(8)$. Since Weyl spinors of $Cl(8)$ are not necessarily pure, the supersymmetric backgrounds considered in \cite{patalong} are not the most general.

One fruitful approach to classifying supersymmetric flux vacua is through the properties of their Killing spinor ansatz. The approach of this paper is in the
same spirit. In the following we will use the term {\it pure background} to refer to a bosonic supersymetric background which is topologically of the  form $\mathbb{R}^{1,p}\times\cm_{2d}$ and for which the internal parts (along $\cm_{2d}$) of the Killing spinors are nowhere-vanishing pure spinors of $Cl(2d)$. We will also assume the internal manifold $\cm_{2d}$ is spin and is equipped with a Riemannian metric.
As will be reviewed in section \ref{sec:pure}, pure backgrounds include the important case of compactifications on Calabi-Yau (CY) manifolds.
Moreover the existence of a metric and a nowhere-vanishing\footnote{As we will see in section \ref{sec:pure}, the condition for the pure spinors to be nowhere-vanishing follows from the requirement that the background admits kappa-symmetric probe branes which do not break the background supersymmetry.} pure spinor reduces the structure group of $\cm_{2d}$ to $SU(d)$. Hence pure backgrounds of type II  supergravities can also be thought of as generalized $SU(d)\times SU(d)$ structure backgrounds.

In \cite{pt} it was shown that the conjecture of \cite{patalong} does not capture the full set of supersymmetry equations: one additional pure-spinor equation needs to be included which does not correspond to a calibration for a D-brane, a result consistent with \cite{toma,mart}. Rather this additional equation corresponds to an analytic continuation of the calibration for instantonic D-branes. This was shown
in \cite{pt} in the case of strict pure backgrounds and was recently proven for  general pure backgrounds in \cite{rosa}.

In the formulation of \cite{pt}, background supersymmetry is given by three pure-spinor equations. The first of those expresses the condition that the internal space is generalized CY in the sense of Hitchin \cite{hitch}. The remaining two can be thought of as exactness equations for the flux and its Hodge-dual: one is given in terms of a twisted differential, while the other is given in terms of a twisted generalized Dolbeualt operator. This is in fact a general pattern that has been observed in  pure\footnote{Since in dimensions $2d\leq6$ all Weyl spinors are pure, for $d=2,3$ pure backgrounds are generic supersymmetric bosonic type II backgrounds.} type II backgrounds of the form $\mathbb{R}^{1,9-2d}\times\cm_{2d}$, $d=2,3,4$:
\begin{itemize}
\item The internal manifold is generalized CY
\item The flux (and/or its Hodge dual) is exact,
\end{itemize}
where the exactness is given in terms of a twisted differential or a twisted generalized Dolbeault operator.

In the present paper we answer the question of whether this general pattern  also holds in the generic case where `spacetime' is a Riemannian manifold $\cm_{10}$. The use of Euclidean signature is necessary if one wants to use the machinery of generalized complex geometry (see however \cite{toma} for a different apporach in Lorentzian signature). We will thus consider pure backgrounds of Euclidean type II supergravities in ten dimensions.
We emphasize that pure backgrounds do not correspond to the most general Killing spinor ansatz -- for which it is known that the simplified patterns described above do not hold in general \cite{toma}.

We show that most but not all of the content of the supersymmetry equations can be rephrased in terms of two pure-spinor equations,
\boxedeq{\spl{\label{main}
\d_H\big(\alpha^2e^{-\phi} \Psi_2\big)&=0\\
i\bar{\partial}^{\mathcal{I}_2}_H\big(e^{-\phi}\mathrm{Im}\Psi_1\big)&=F^-
~,}}
where the notation will be explained in detail in the following, while the remaining supersymmetry equations cannot be put in a similar form.

By restricting to real Euclidean type IIA pure backgrounds
the results of this paper uplift to the `physical' Lorentzian eleven-dimensional supergravity. In \cite{h} it was proposed that `M-theory on a timelike circle' could be thought of as the strong-coupling limit of  real Euclidean IIA string theory in ten dimensions. The idea of a Euclidean theory growing a time direction at strong coupling was also taken up more recently in \cite{hl} in the context of five-dimensional Euclidean super Yang-Mills.

The outline of the remainder of the paper is as follows. In section \ref{sec:pure} we
explain the general setup and the properties of Euclidean type II
supersymmetric backgrounds. To show that (\ref{main}) do not capture the full content of supersymmetry, it suffices to work with a so-called `strict' Killing-spinor ansatz. The analysis of the supersymmetry conditions in terms
of strict $SU(5)$ structures is given in section \ref{sec:su5}. These conditions
are then reformulated in the language of generalized complex geometry in section \ref{sec:susygen}, where we show that most but not all of the supersymmetry
conditions are contained in (\ref{main}). In section \ref{sec:dynamic} we go
beyond the strict Killing-spinor ansatz and show that the first line of
(\ref{main}) holds in the case of general $SU(5)\times SU(5)$ structure pure backgrounds. In section \ref{sec:uplift} we uplift the solutions of real IIA
Euclidean supergravity to real Lorentzian eleven-dimensional supergravity. Section \ref{sec:int} discusses the integrability of the supersymmetry conditions, i.e. the conditions under which all supergravity equations of motion are satisfied. In order to illustrate the formalism we give in section \ref{sec:example} a simple class of examples of eleven-dimensional supersymmetric backgrounds with non-trivial four-form which satisfy all equations of motion.

\section{Pure backgrounds of Euclidean type II }\label{sec:pure}

As was shown in \cite{berg2},
all variant type II
supergravities in ten dimensions can be obtained by `holomorphic complexification' of the standard type II supergravities, whereby one
complexifies all fields appearing in the action in such a way that no complex conjugates appear. This is completely straightforward for the bosonic fields; for the fermions one must first replace all Dirac conjugates (${\psi}^\dagger\gamma^0$) that appear in the action by Majorana conjugates (${\psi}^TC^{-1}$). The same holomorphic complexification must also be applied
to the supersymmetry transformations. This procedure then guarantees that the resulting complexified action remains invariant under the complexified supersymmetry variations.

To obtain the type II theories in Euclidean signature one may  work with flat gamma matrices that satisfy a Lorentzian Clifford algebra, and introduce a vielbein with an imaginary time-like component. Alternatively one may work with flat gamma matrices that satisfy a Euclidean Clifford algebra and with real vielbeine; this is the approach we will adopt here. As explained in \cite{berg2}, the reason why one can do this is that gamma matrices always appear in the combination $\gamma^m=e^m{}_a\gamma^a$ so that simultaneously Wick-rotating the flat gamma matrices $\gamma^a\rightarrow i\gamma^a$ and the vielbein $e_m{}^a\rightarrow i e_m{}^a$ leaves the curved gamma matrices $\gamma^m$ invariant. Note that this implies that the chirality matrix $\gamma_{11}$ should also be Wick-rotated to the Euclidean-signature chirality matrix, see appendix \ref{app1} for our conventions.

We will consider Euclidean type II theories on Riemannian spin manifolds $\cm_{10}$. We will adopt a `democratic' formulation whereby one doubles the Ramond-Ramond (RR) field-strengths $F$ while simultaneously imposing an imaginary twisted self-duality condition:\footnote{We use polyform notation for the RR fields. Our conventions for type II supergravity are obtained from \cite{march}, see appendix A therein, by holomorphically complexifying and imposing Euclidean signature. Condition (\ref{sd}) can be obtained by requiring that when applied to the supersymmetry variations, in order to eliminate the
redundant RR fieldstrengths, it gives back the standard (non-democratic) variations. The sign in (\ref{sd}) is thus correlated with the choice of sign in
the gamma-matrix Hodge duality condition (\ref{eq:gamhodge}).}
\eq{\label{sd}F=-i\star\sigma(F)~.}
The supersymmetry parameters of complexified type II theories are expressed in terms of two complex Weyl spinors of $Spin(10)$, $\epsilon_1$, $\epsilon_2$,
so that:
\eq{\gamma_{11}\epsilon_1=\epsilon_1~;~~~
\gamma_{11}\epsilon_2=\left\{
\begin{array}{rl}
-\epsilon_2 &, ~~~\mathrm{in~~IIA}\nn\\
\epsilon_2 &, ~~~\mathrm{in~~IIB}
\end{array} \right.
~.}
We will consider pure backgrounds, in other words we are looking
for bosonic supersymmetric backgrounds for which the Killing spinors $\epsilon_1$, $\epsilon_2$ are pure spinors.
We will also demand that $\epsilon_1$, $\epsilon_2$ have equal norms,
\eq{\label{norm}
|\epsilon_1|^2=|\epsilon_2|^2~,
}
in analogy with Lorentzian backgrounds where this condition follows from the requirement that the background admits supersymmetry-preserving kappa-symmetric probe branes.

In these conventions for both type II supergravities the Killing spinor equations for a bosonic background are given by:
\eq{\spl{\label{kse}
\delta \lambda^1 &= \big(\underline{\p}\phi + \frac{1}{2} \underline{H} \big) \epsilon_1+  \frac{1}{16} e^{\phi} \g^m \underline{F}\g_m \g_{11} ~\!\epsilon_2 = 0 \\
\delta \lambda^2 &= \big( \underline{\p }\phi - \frac{1}{2} \underline{H} \big) \epsilon_2 -  \frac{1}{16} e^{\phi} \g^m \sigma(\underline{F})\g_m \g_{11} ~\! \epsilon_1 = 0 \\
\delta \psi^1_m &= \big( \nabla_m + \frac{1}{4} \underline{H}_m \big) \epsilon_1 + \frac{1}{16} e^{\phi} \underline{F}\g_m \g_{11}~\! \epsilon_2 = 0 \\
\delta \psi^2_m &= \big( \nabla_m - \frac{1}{4} \underline{H}_m \big) \epsilon_2 -\frac{1}{16} e^{\phi} \sigma(\underline{F})\g_m \g_{11}~\! \epsilon_1 = 0 \;,
}}
where for any $(p+q)$-form $S$ we define:
\eq{
\underline{S}_{m_1\dots m_q}\equiv\frac{1}{p!}\gamma^{n_1\dots n_p}
{S}_{n_1\dots n_pm_1\dots m_q}
~.}
Note that the last two lines in (\ref{kse}) can be viewed as a system of linear first-order differential equations for $X\equiv(\epsilon_1,\epsilon_2)$. It follows that (under certain smoothness assumptions for the coefficients which are given in terms of the background bosonic fields) if $X$ vanishes at a point it should be identically zero. Given the norm condition (\ref{norm}) we see that $\epsilon_1$, $\epsilon_2$ must both be nowhere-vanishing.

The existence of a nowhere-vanishing pure Weyl spinor on the Riemannian spin manifold $\cm_{10}$ implies the reduction of the structure group from $SO(10)$ to $SU(5)$. A pure Weyl spinor may be defined as a spinor which is annihilated by those gamma matrices that are antiholomorphic (or holomorphic, depending on the convention) with respect to some almost complex structure\footnote{An equivalent definition for a pure spinor $\eta$ in $2d$ dimensions is that $\tilde{\eta}\g_{m_1\dots m_p}\eta=0$, for $p<d$.}. Hence there is a correspondence between line bundles of pure spinors and almost complex structures on $\cm_{10}$. As is well-known, the existence of an almost complex structure reduces the structure group from $SO(10)$ to $U(5)$. Demanding that the line bundle of pure spinors should have a global section (i.e. there is a nowhere-vanishing pure spinor) the structure group is further reduced to $SU(5)$.

It can be seen that CY fivefolds, which are manifolds of $SU(5)$ holonomy, are special cases of pure backgrounds in ten dimensions. Indeed, a CY $d$-fold possesses a covariantly-constant Weyl spinor $\eta$ of $Spin(2d)$ which is annihilated by all antiholomorphic gamma matrices, i.e. $\eta$ is pure.
The fact that for a CY of full $SU(5)$ holonomy (and not a proper subgroup thereof) the covariantly constant spinor $\eta$ must be pure can be seen directly as follows: if $\eta$ were not pure the complex vector field $K^m$ given by the spinor bilinear $K^m\equiv\tilde{\eta}\gamma^m\eta$ would be  covariantly constant, nowhere-vanishing and invariant under the action of $SU(5)$. Since the CY admits an integrable almost complex structure with respect to which $K^m$ is holomorphic, the fibers of the tangent bundle are isomorphic to $\mathbbm{C}^5$ and the action of $SU(5)$ on the vector field is the fundamental action. However since the only trivial orbit of $SU(5)$ is the zero vector, $K^m$ cannot be both nowhere-vanishing and invariant under the reduced structure group unless the structure group admits a further reduction to $SU(4)$, leading to a contradiction.


\section{Supersymmetry in terms of $SU(5)$ structures}\label{sec:su5}

As follows from the discussion in section \ref{sec:pure} each one of the two pure nowhere-vanishing spinors $\epsilon_1$, $\epsilon_2$ defines an $SU(5)$ structure on $\cm_{10}$. When the two spinors are independent this is sometimes called a dynamic $SU(5)\times SU(5)$ structure.
In this section we will make the simplifying assumption that $\epsilon_1$, $\epsilon_2$ are not independent but rather that $\epsilon_2$ is proportional to either $\epsilon_1$ (in IIB) or to $\epsilon_1^c$ (in IIA), see appendix  \ref{app1} for our definition of the spinor complex conjugate. We will return  to the general case of dynamic $SU(5)\times SU(5)$ structure in section \ref{sec:dynamic}.

\subsection{Euclidean IIA}\label{sec:eiia}

Our strict $SU(5)$ ansatz reads:
\eq{\label{spinora}\epsilon_1= \alpha\eta~;~~~\epsilon_2=\alpha e^{i\theta}\eta^c~,}
where $\eta$ is a pure, positive-chirality spinor of unit norm; the scalar function $\alpha$ can be taken to be real without loss of generality; $\theta$ is a phase which will be position-dependent in general. Note that $\eta^c$ has negative chirality, since the irreducible (sixteen-dimensional) spinor representation of $Spin(10)$ is complex. Inserting the spinor ansatz (\ref{spinora}) into (\ref{kse}) using the $SU(5)$ tensor decompositions of section \ref{sec:tensor} and equation (\ref{tornabl}) in order to express the covariant derivative in terms of $SU(5)$ torsion classes, the Killing spinor equations
reduce to the following set of algebraic relations:
\eq{\label{iiasol} \spl{
\t  &= \mathrm{const}\\
\a&=\mathrm{const}\times e^{\frac12\phi}\\
W_1 &= -i h^{(2,0)}  \\
W_2 &= 4 i e^{i\t} e^{\phi} f_4^{(3,1)} \\
W_3 &= -i h^{(2,1)} \\
W_4 &=  -i h^{(1,0)} \\
W_5 &= \p^+ \phi   \\
h^{(2,0)} &= \frac{3}{16}e^\phi  e^{i \t} \left( f_{2}^{(2,0)} - if_{4}^{(2,0)} \right)\\
f_0 &= 3i f_2 + 4 f_4 \\
f_2^{(1,1)} &=  3i f_4^{(1,1)} \\
f_4^{(1,0)} &= - \frac{1}{2} e^{- \phi} e^{i\t} \left( \p^+ \log\a + i h^{(1,0)} \right)~,
}}
so that the parameterization of the solution can be given in terms the real constant $\theta$, the complex scalars $\phi$, $f_2$, $f_4$, and the forms $h^{(1,0)}$, $f_{2}^{(2,0)}$,  $f_{4}^{(2,0)}$, $h^{(2,1)}$, $f_{4}^{(3,1)}$;  we use the symbols $\partial^{+}$, $\p^-$ to denote the projection of the exterior differential onto the (1,0), (0,1) parts respectively; in addition we have the following reality conditions:
\eq{\spl{\label{realiiasol}
\big(f_{2}^{(2,0)}  - if_{4}^{(2,0)}  \big)^* &= - \big(  f_{2}^{(0,2)} - i f_{2}^{(0,2)}  \big)~;~~~\big({f}_{4}^{(1,0)}\big)^*  = {f}_{4}^{(0,1)} ~;~~~\big( f_4^{(3,1)} \big)^* =  f_4^{(1,3)} \\
\big( h^{(1,0)} \big)^* &= - h^{(0,1)}~;~~~ \big({h}^{(2,0)} \big)^* = - {h}^{(0,2)} ~;~~~\big( h^{(2,1)} \big)^* = - h^{(1,2)}~.
}}

\subsubsection*{Real Euclidean IIA}

The above solution and in particular the conditions (\ref{realiiasol}) are consistent with, but less stringent than, the reality conditions of the real Euclidean IIA theory. Indeed in the real Euclidean IIA theory we have, adapting  \cite{berg2} to our conventions:
\eq{\label{realiia}H^*=-H~;~~~F^*=\sigma(F)~,}
with all remaining bosonic fields real.
Moreover in the real Euclidean IIA theory the Killing spinor parameters are constrained to satisfy the following reality condition:
\eq{\label{realiiaspin}\epsilon_2=-i\epsilon_1^c~.}
This is also consistent with the spinor ansatz (\ref{spinora}) and the solution (\ref{iiasol}) provided \eq{\label{realiiat}\theta=-\frac{\pi}{2}~.}
\subsection{Euclidean IIB}\label{sec:eiib}
In this case the strict $SU(5)$ ansatz reads:
\eq{\label{spinorb}\epsilon_1= \alpha\eta~;~~~\epsilon_2=\alpha e^{i\theta}\eta~,}
where as in the case of IIA $\eta$ is a pure, positive-chirality spinor of unit norm; the scalar function $\alpha$ can be taken to be real without loss of generality; $\theta$ is a phase which will be position-dependent in general.  Inserting the spinor ansatz (\ref{spinorb}) into (\ref{kse}) using the $SU(5)$ tensor decompositions of section \ref{sec:tensor} and equation (\ref{tornabl}) in order to express the covariant derivative in terms of $SU(5)$ torsion classes, the Killing spinor equations reduce to the following set of algebraic relations:
\eq{\spl{\label{iibsol}
W_1&=0\\
W_2&=0\\
W_3^*&= -i e^{\phi} \big( \cos\t f_3^{(1,2)} - \sin\t f_5^{(1,2)} \big)\\
W_4^* &= \frac{1}{2} \p^- \phi \\
W_5^* &= \p^-\big( \phi - 2 \log \a - i \t \big)\\
h^{(0,1)} &= \frac{1}{2} \p^- \t  \\
{h}^{(2,0)} &=  0\\
h^{(1,2)} &= -i e^{\phi}  ( \sin\t f_3^{(1,2)} + \cos\t f_5^{(1,2)}) \\
f_{1}^{(0,1)} &= i \p^- \big(e^{-\phi} \sin\t \big) \\
f_{3}^{(0,1)} &= \frac{i}{2} \p^- \big(e^{-\phi} \cos\t \big) \\
{f}_{3}^{(2,0)} &=0\\
f_5^{(1,4)} &= 0\\
e^\phi e^{i\t} \big( \frac{1}{4} f_{1}^{(1,0)} + i f_{3}^{(1,0)} - \frac{3}{2} f_{5}^{(1,0)} \big)
&= -{\p^+}\big( 2 \log \a  -\frac{i}{2}\t \big)- i h^{(1,0)} \\
e^\phi e^{- i\t} \big(- \frac{1}{4} f_{1}^{(1,0)} + i f_{3}^{(1,0)} + \frac{3}{2} f_{5}^{(1,0)} \big)
&=  -{\p^+}\big( 2 \log \a  +\frac{i}{2}\t \big)+ i h^{(1,0)} ~,
}}
so that the real scalars $\alpha$, $\t$, the complex scalar $\phi$ and the complex forms $f_3^{(1,2)}$,  $f_5^{(1,2)}$ can be thought of as free `parameters' of the solution. It follows from the vanishing of the first two torsion classes (the first two lines above) that $\cm_{10}$ is complex. In this case we may therefore introduce complex coordinates and identify $\p^+$, $\p^-$ with the holomorphic, antiholomorphic differential respectively.
\\
\\
As noted in \cite{berg2}, unlike IIA, there is no real version of Euclidean type IIB supergravity.

\section{Supersymmetry in terms of generalized geometry}\label{sec:susygen}

For a brief summary of the elements of generalized geometry that we will use here we refer the reader to appendix C.1 of \cite{pt}. A pedagogical introduction to generalized geometry for physicists was given in \cite{koer}.

{
As already mentioned, the backgrounds we are considering here admit two nowhere-vanishing, pure Killing spinors $\epsilon_1$, $\epsilon_2$.
We define
\eq{\label{puregen}\underline{\Psi}_1 = -\frac{2^5}{|\epsilon|^2}~\! \epsilon_1 \otimes \widetilde{\epsilon_2^c}~;~~~
\underline{\Psi}_2 =  -\frac{2^5}{|\epsilon|^2} ~\!\epsilon_1 \otimes \widetilde{\epsilon_2}~,}

where $|\epsilon|^2\equiv|\epsilon_1|^2=|\epsilon_2|^2$ is the norm of the Killing spinors. Using the Fierz identity
\al{
\chi_1 \otimes \tilde{\chi}_2 = \frac{1}{2^5} \sum_{p=0}^{10} \frac{1}{p!} \tilde{\chi}_2 \g_{m_1 ...m_p} \chi_1 \g^{m_p ... m_1}
}
and, after a choice of volume form, the Clifford map \eqref{cliff}, one can identify these bispinors with polyforms on $\cm_{10}$. In the language of generalized complex geometry, there is a natural action of the generalized tangent bundle $T \oplus T^*$ on $\bigwedge\nolimits^{\!\bullet} T^*$. This action obeys the Clifford algebra with respect to the natural metric of signature $(10,10)$ on $T\oplus T^*$, which pairs one-forms and vector fields with one another.
Thus $\Psi_{1,2}$ can be viewed as spinors of $Cl(10,10)$ which are pure by virtue of the purity of $\e_{1,2}$. Just as there is a correspondence between almost complex structures on $\cm_{10}$ and line bundles of pure Weyl spinors of $Cl(10)$, there is a correspondence between generalized almost complex structures on $T\oplus T^*$ and line bundles of pure spinors of $Cl(10,10)$. Hence the $\Psi_{1,2}$ can be identified with two generalized
almost complex structures which are compatible (i.e. they commute).
Since  $\Psi_{1,2}$ are nowhere-vanishing, the structure group of the generalized tangent bundle reduces from $O(10,10)$ to $SU(5) \otimes SU(5)$.
}
%
%

In the following we will denote by $\mathcal{I}_1$, $\mathcal{I}_2$ the generalized almost complex structure whose  $+i$ eigenbundle is isomorphic to the space of generalized gamma matrices annihilating $\Psi_{1}$, $\Psi_2$ respectively. Moreover polyforms/bispinors admit a double decomposition in terms of the eigenspaces $U^{(k,l)}$ of $(\mathcal{I}_1$, $\mathcal{I}_2)$ with eigenvalue $(k,l)$. Specifically in ten dimensions we may decompose a polyform $\Phi$:
\eq{\label{polydec}\Phi=\sum_{k=-5}^5\left(\Phi^{(k,|k|-5)}+
\Phi^{(k,|k|-5+2)}+\dots+\Phi^{(k,5-|k|)}\right)
~,}
such that $\Phi^{(k,l)}\in U^{(k,l)}$. This leads to the {\it generalized Hodge diamond}:
\eq{\label{hodge1}
\begin{array}{c}\vspace{.1cm}
U^{(0,5)}\\
U^{(1,4)}\hspace{1cm}
U^{(-1,4)}\\
U^{(2,3)}\hspace{1cm}U^{(0,3)}\hspace{1cm}
U^{(-2,3)}\\
U^{(3,2)}\hspace{1cm}U^{(1,2)}\hspace{1cm}
U^{(-1,2)}\hspace{1cm}U^{(-3,2)}\\
U^{(4,1)}\hspace{1cm}U^{(2,1)}\hspace{1cm}
U^{(0,1)}\hspace{1cm}
U^{(-2,1)}\hspace{1cm}U^{(-4,1)}\\
U^{(5,0)}\hspace{1cm}U^{(3,0)}\hspace{1cm}
U^{(1,0)}\hspace{1cm}
U^{(-1,0)}\hspace{1cm}U^{(-3,0)}
\hspace{1cm}U^{(-5,0)}\\
U^{(4,-1)}\hspace{1cm}U^{(2,-1)}\hspace{1cm}
U^{(0,-1)}\hspace{1cm}
U^{(-2,-1)}\hspace{1cm}U^{(-4,-1)}\\
U^{(3,-2)}\hspace{1cm}U^{(1,-2)}\hspace{1cm}
U^{(-1,-2)}\hspace{1cm}U^{(-3,-2)}\\
U^{(2,-3)}\hspace{1cm}U^{(0,-3)}\hspace{1cm}
U^{(-2,-3)}\\
U^{(1,-4)}\hspace{1cm}
U^{(-1,-4)}\\
U^{(0,-5)}
\end{array}
}
It will be useful to introduce the {\it Mukai pairing} which is an inner product on the space of polyforms, see appendix \ref{sec:mukai} for the exact definition.
The terms in the decomposition (\ref{polydec}) can then be expressed (up to normalization) in terms of the Mukai paring:
\eq{\label{proj}\Phi^{(k,l)}\propto\langle u^{(-k,-l)},\Phi\rangle~\! u^{(k,l)} ~,}
where $u^{(k,l)}$ is an orthogonal basis for $U^{(k,l)}$, so that
\eq{\langle u^{(k,l)},u^{(p,q)}\rangle\propto\delta_{k+p}\delta_{l+q}~.}
As a corollary we note that, a polyform vanishes if and only if it has vanishing Mukai pairings with all the forms in the Hodge diamond. We will make use of this observation in section \ref{iiagen}.

In ten dimensions an explicit orthogonal basis
for the Hodge diamond is given by
\eq{\label{hodge2}
\begin{array}{c}\vspace{.1cm}
\underline{\Psi}_1\gamma_{5-}\\
\underline{\Psi}_1\gamma_{4-}\hspace{1cm}
\gamma_{4-}\underline{\Psi}_1^*\\
\underline{\Psi}_1\gamma_{3-}\hspace{1cm}\gamma_{1+}\underline{\Psi}_1\gamma_{4-}\hspace{1cm}
\gamma_{3-}\underline{\Psi}_1^*\\
\underline{\Psi}_1\gamma_{2-}\hspace{1cm}\gamma_{1+}\underline{\Psi}_1\gamma_{3-}\hspace{1cm}
\gamma_{3-}\underline{\Psi}_1^*\gamma_{1+}\hspace{1cm}\gamma_{2-}\underline{\Psi}_1^*\\
\underline{\Psi}_1\gamma_{1-}\hspace{1cm}\gamma_{1+}\underline{\Psi}_1\gamma_{2-}\hspace{1cm}
\gamma_{2+}\underline{\Psi}_1\gamma_{3-}\hspace{1cm}
\gamma_{2-}\underline{\Psi}_1^*\gamma_{1+}\hspace{1cm}\gamma_{1-}\underline{\Psi}_1^*\\
\underline{\Psi}_1\hspace{1cm}\gamma_{1+}\underline{\Psi}_1\gamma_{1-}\hspace{1cm}
\gamma_{2+}\underline{\Psi}_1\gamma_{2-}\hspace{1cm}
\gamma_{2-}\underline{\Psi}_1^*\gamma_{2+}\hspace{1cm}\gamma_{1-}\underline{\Psi}_1^*\gamma_{1+}
\hspace{1cm}\underline{\Psi}_1^*\\
\gamma_{1+}\underline{\Psi}_1\hspace{1cm}\gamma_{2+}\underline{\Psi}_1\gamma_{1-}\hspace{1cm}
\gamma_{3+}\underline{\Psi}_1\gamma_{2-}\hspace{1cm}
\gamma_{1-}\underline{\Psi}_1^*\gamma_{2+}\hspace{1cm}\underline{\Psi}_1^*\gamma_{1+}\\
\gamma_{2+}\underline{\Psi}_1\hspace{1cm}\gamma_{3+}\underline{\Psi}_1\gamma_{1-}\hspace{1cm}
\gamma_{1-}\underline{\Psi}_1^*\gamma_{3+}\hspace{1cm}\underline{\Psi}_1^*\gamma_{2+}\\
\gamma_{3+}\underline{\Psi}_1\hspace{1cm}\gamma_{4+}\underline{\Psi}_1\gamma_{1-}\hspace{1cm}
\underline{\Psi}_1^*\gamma_{3+}\\
\gamma_{4+}\underline{\Psi}_1\hspace{1cm}
\underline{\Psi}_1^*\gamma_{4+}\\
\gamma_{5+}\underline{\Psi}_1
\end{array}
}
where a gamma-matrix $\gamma_+$, $\gamma_-$ acting on the left is understood as being holomorphic, respectively antiholomorphic, with respect to the ordinary (not generalized) almost complex structure associated with the pure spinor $\epsilon_1$; a gamma-matrix $\gamma_+$, $\gamma_-$ acting on the right is understood as being holomorphic, respectively antiholomorphic, with respect to the ordinary almost complex structure associated with the pure spinor $\epsilon_2$; the notation
$\gamma_{p\pm}$ stands for the product $\gamma_{a_1\pm}\dots\gamma_{a_p\pm}$ of $p$ holomorphic or antiholomorphic gamma-matrices.

In the following two subsections we will show that most (but not all) of the supersymmetry equations for both Euclidean type II theories can be cast in the form of the two polyform equations (\ref{main}),\footnote{It is interesting to note that, if $\Psi_{1,2}$ are thought of as spinors of $Cl(10,10)$, the first line in (\ref{main}) can be understood as a Dirac equation for  $\Psi_{1}$ \cite{waldram,andriot, hohm1, hohm2}; the second line in (\ref{main}) can also be given a
similar interpretation. We would like to thank to D.~Andriot for bringing this to our attention.}
where $F^-$ is the projection of $F$ onto the subspace $\sum_k\sum_{l < 0}\oplus ~\!U^{(k,l)}$. The generalized twisted Dolbeault operator $\partial_H^{\mathcal{I}_2}$ is defined as follows. The first line of (\ref{main}) expresses the fact that the manifold $\cm_{10}$ is generalized CY; in particular it is generalized complex, i.e. the almost complex structure $\mathcal{I}_2$ (associated with $\Psi_2$) is integrable. This implies that $\d_H$ maps a polyform $\Phi^{(l)}$ with $\mathcal{I}_2$-eigenvalue $+il$ to the subspaces of polyforms with $\mathcal{I}_2$-eigenvalues  $+i(l+1)$, $+i(l-1)$:
\eq{
\d_H (\Phi^{(l)} )=\left(\d_H\Phi\right)^{(l+1)}
+\left(\d_H\Phi\right)^{(l-1)}
~.}
We can thus define a twisted {\it generalized Dolbeault operator} $\partial_H^{\mathcal{I}_2}$ associated with the integrable almost complex structure $\mathcal{I}_2$ via
\eq{\label{dolbeault}\partial_H^{\mathcal{I}_2}\Phi\equiv\left(\d_H\Phi\right)^{(l+1)}~,
~~~~~\bar{\partial}_H^{\mathcal{I}_2}\Phi\equiv\left(\d_H\Phi\right)^{(l-1)}~.}
In particular $\bar{\partial}_H^{\mathcal{I}_2}\big(e^{-\phi}\mathrm{Im}\Psi_1\big)$ which appears in (\ref{main}) is a polyform with $\mathcal{I}_2$-eigenvalue $-i$. This is because $\Psi_1$ has zero $\mathcal{I}_2$-eigenvalue, as can be seen from the explicit basis of the Hodge-diamond given above.

Let us also note that equation (1.2) of \cite{pt}\footnote{For the convenience of the reader let us recall that equation (1.2) of \cite{pt} reads:
$$\d^{\mathcal{I}_2}_H\big(e^{-\phi}\mathrm{Im}\Psi_1\big)=F~,$$ where $\d^{\mathcal{I}_2}_H\equiv i(\bar{\partial}_H^{\mathcal{I}_2}
-\partial_H^{\mathcal{I}_2})$.} can be written equivalently in the form of the second line of (\ref{main}), in terms of the Dolbeault operator $\bar{\partial}^{\mathcal{I}_2}_H$. However the converse is not true: the second line of  (\ref{main}) cannot be put in the form of  equation (1.2) of \cite{pt}. This is because the fluxes $H$, $F$ are not real, hence
$F^- +(F^-)^*\neq F$ and $(\bar{\partial}^{\mathcal{I}_2}_H)^* \neq \partial^{\mathcal{I}_2}_H$.

\subsection{Euclidean IIA}\label{iiagen}

Inserting the strict $SU(5)$ spinor ansatz (\ref{spinora}) in (\ref{puregen}) and using the Fierz identity and Clifford map,  we find
\eq{\label{bla}\Psi_1 = e^{-i\t}\Omega~;~~~
\Psi_2 =  -e^{i\t}e^{-iJ}~,}
where we have taken into account the definition (\ref{su5}) of the
$SU(5)$ structure.
From (\ref{torj}) we can then calculate $\d_H\Psi_2$ and
$\d_H\mathrm{Im}\Psi_1$, which appear on the left-hand side of the equations in (\ref{main}), in terms of the torsion classes. Moreover to calculate the second line of (\ref{main}) we need to project onto the subspace $\sum_k\sum_{l < 0}\oplus ~\!U^{(k,l)}$. One way to do this is by calculating
the Mukai pairings of both sides of the equation with $u^{(p,q)}$ for $q>0$, cf. (\ref{proj}). Specializing to the case of a strict $SU(5)$ structure, taking (\ref{bla}) into account, the explicit basis $\{u^{(k,l)}\}$ for the Hodge diamond can be represented as:
\eq{\label{hodge3}
\begin{array}{c}\vspace{.1cm}
\underline{\Omega}\gamma_{5+}\\
\underline{\Omega}\gamma_{4+}\hspace{1cm}
\gamma_{4-}\underline{\Omega}^*\\
\underline{\Omega}\gamma_{3+}\hspace{1cm}\gamma_{1+}\underline{\Omega}\gamma_{4+}\hspace{1cm}
\gamma_{3-}\underline{\Omega}^*\\
\underline{\Omega}\gamma_{2+}\hspace{1cm}\gamma_{1+}\underline{\Omega}\gamma_{3+}\hspace{1cm}
\gamma_{3-}\underline{\Omega}^*\gamma_{1-}\hspace{1cm}\gamma_{2-}\underline{\Omega}^*\\
\underline{\Omega}\gamma_{1+}\hspace{1cm}\gamma_{1+}\underline{\Omega}\gamma_{2+}\hspace{1cm}
\gamma_{2+}\underline{\Omega}\gamma_{3+}\hspace{1cm}
\gamma_{2-}\underline{\Omega}^*\gamma_{1-}\hspace{1cm}\gamma_{1-}\underline{\Omega}^*\\
\underline{\Omega}\hspace{1cm}\gamma_{1+}\underline{\Omega}\gamma_{1+}\hspace{1cm}
\gamma_{2+}\underline{\Omega}\gamma_{2+}\hspace{1cm}
\gamma_{2-}\underline{\Omega}^*\gamma_{2-}\hspace{1cm}\gamma_{1-}\underline{\Omega}^*\gamma_{1-}
\hspace{1cm}\underline{\Omega}^*\\
\gamma_{1+}\underline{\Omega}\hspace{1cm}\gamma_{2+}\underline{\Omega}\gamma_{1+}\hspace{1cm}
\gamma_{3+}\underline{\Omega}\gamma_{2+}\hspace{1cm}
\gamma_{1-}\underline{\Omega}^*\gamma_{2-}\hspace{1cm}\underline{\Omega}^*\gamma_{1-}\\
\gamma_{2+}\underline{\Omega}\hspace{1cm}\gamma_{3+}\underline{\Omega}\gamma_{1+}\hspace{1cm}
\gamma_{1-}\underline{\Omega}^*\gamma_{3-}\hspace{1cm}\underline{\Omega}^*\gamma_{2-}\\
\gamma_{3+}\underline{\Omega}\hspace{1cm}\gamma_{4+}\underline{\Omega}\gamma_{1+}\hspace{1cm}
\underline{\Omega}^*\gamma_{3-}\\
\gamma_{4+}\underline{\Omega}\hspace{1cm}
\underline{\Omega}^*\gamma_{4-}\\
\gamma_{5+}\underline{\Omega}
\end{array}
}
As in (\ref{hodge2}), the lines above correspond to fixed $l$, varying from $l=5$ for the top line to $l=-5$ for the bottom line; the columns correspond to fixed $k$, varying from $k=5$ for the leftmost column to $k=-5$ for the rightmost column.
For example we have: $u^{(2,-1)}_{ab,c}=\gamma_a^+\gamma_b^+\underline{\Omega}\gamma^+_c$,
 $u^{(-2,-1)}_{ab,c}=\gamma^-_c\underline{\Omega}^*\gamma_a^-\gamma_b^-$, etc.
Note that since in the case of IIA with strict $SU(5)$ structure the almost complex structures associated with $\epsilon_{1}$ and $\epsilon_{2}^c$ are identical, it is not necessary to distinguish between $\gamma^{\pm}$ acting on the left or on the right of $\underline{\Psi}_1\sim\underline{\Omega}$.

With the help of the formul\ae{} in section \ref{sec:mukai} we then see that equations (\ref{main}) are equivalent to the susy equations in (\ref{iiasol}) except for the last line therein. Moreover the reality conditions (\ref{realiiasol}) and the condition $\theta=\mathrm{const}$ are also not captured by (\ref{main}); however as we explained in section \ref{sec:eiia} they follow from imposing the reality conditions of Euclidean IIA.

{
Let us recapitulate: the derivation above consists in taking a double decomposition of polyforms $U^{(k,l)}$ with respect to $(\mathcal{I}_1,\mathcal{I}_2)$, projecting onto each $U^{(k,l)}$ using the Mukai pairing, then summing over all $k$ since the Dolbeault operator in \eqref{main} is defined using only $\mathcal{I}_2$. An alternative derivation is as follows: one decomposes the polyforms only with respect to $\mathcal{I}_2$, projects onto the corresponding eigenspaces $U^{(l)}$ of $+il$ eigenvalue,  and then sums over $l < 0$. Thus, we need to know what $U^{(l)}$ is for a generalized complex structure induced by a symplectic form. Fortunately, this is known: it is given by \cite{cavalcanti}
\al{
U^{(l)} = \{ e^{-i J} e^{\frac{1}{2} i \L} \Phi    \;|\; \Phi \in \bigwedge\nolimits^{\!n-l} T^*  \}\;,
}
where $n=5$ in our case and $\L : \bigwedge\nolimits^{\!k} T^* \rightarrow \bigwedge\nolimits^{\!k-2} T^*$ is defined as\footnote{Our sign difference with respect to \cite{cavalcanti} comes from the fact that our action of $\mathcal{I}_2$ and $\L$ on polyforms differ by a sign; our convention is more natural from the point of view of index contractions.}
\all{
\L ( \Phi) = J \lrcorner \Phi \;.
}
We can thus decompose each polyform $\Phi = \sum_{k=0}^{10} \Phi_k$, $\Phi_k \in \bigwedge\nolimits^{\!k} T^*$ to conclude that
\al{
\Pi^- \left(\Phi\right) = \sum_{l<0} e^{-i J} e^{ \frac{1}{2} i \L} \left(e^{- \frac{1}{2} i \L} e^{i J} \Phi\right)_{5-l} \;.
}
However there is no need to do the decomposition explicitly, since it can be
seen that $\Pi^-(\Phi) = 0$ implies $\left(e^{i J} \Phi\right)_{5-l} =0$, for all $l < 0$. Taking $\Phi = d^{\mathcal{I}_2}_H \left( e^{-\phi} \text{Im} \Psi_1\right) - F $ as in \eqref{main} then yields the desired result.
}

In conclusion, the polyform equations (\ref{main}) come very close to capturing
all of the content of supersymmetry. They do not include, however, the equation in the last line of (\ref{iiasol}).

\subsection{Euclidean IIB}\label{iibgen}

Inserting the strict $SU(5)$ spinor ansatz (\ref{spinorb}) in (\ref{puregen}) we find
\eq{\label{blab}\Psi_1 = - e^{-i\t}e^{-iJ}~;~~~
\Psi_2 =-e^{i\t}\Omega ~,}
where we have taken into account the definition (\ref{su5}) of the
$SU(5)$ structure. From (\ref{torj}) we can then calculate $\d_H\Psi_2$ and
$\d_H \mathrm{Im}\Psi_1$, which appear on the left-hand side of the equations in (\ref{main}), in terms of the torsion classes. Moreover to calculate the second line of (\ref{main}) we need to project onto the subspace $\sum_k\sum_{l < 0}\oplus ~\!U^{(k,l)}$. This projection is much simpler to implement in the case of IIB than in IIA: indeed, as already noticed in section \ref{sec:eiib}, in the case of strict IIB
the manifold $\cm_{10}$ is complex by virtue of supersymmetry and the projection amounts to reducing to the space of polyforms that are sums of $(p,q)$-forms with $p<q$. Moreover the generalized twisted Dolbeault operator $\partial^{\mathcal{I}_2}_H$ reduces to  $\partial_H=\partial+H^{(2,1)}\wedge$, where $\partial$ is the Dolbeault operator on $\cm_{10}$ and $H^{(2,1)}$ is the $(2,1)$-form part of $H$ with respect to the ordinary (not generalized) complex structure of $\cm_{10}$ \cite{pt}. It is then straightforward to see that equations (\ref{main}) are equivalent to the system (\ref{iibsol}) of supersymmetry equations except for the last two equations therein.

\subsection{General proof}\label{sec:dynamic}

The fact that the two generalized equations (\ref{main}) are `mirror-symmetric',  in the sense that their form is the same both in IIA and in IIB, is strong evidence that they should be valid beyond the special case of the strict ansatz. We will show that this is indeed the case for the first of the two equations in (\ref{main}). We hope to return to the general proof of the second equation in the future.

The proof we give here is very similar the one in \cite{scan}, see appendix A therein. The main observation is that for any polyform $\Phi$ we have the following correspondences  under the Clifford map (\ref{cliff}):
\eq{\label{corr}\d x^m\wedge\Phi+\p_m\lrcorner\Phi
\leftrightarrow \gamma^m\underline{\Phi}~;~~~
\d x^m\wedge\Phi-\p_m\lrcorner\Phi
\leftrightarrow (-)^{|\Phi|}\underline{\Phi}\gamma^m
~.}
As a corollary of the above correspondences it is straightforward to show that
for any polyform  $\Phi$ we have
\eq{\spl{\label{int1}
2C\wedge\Phi&\leftrightarrow
\underline{C}{~\!}\underline{\Phi}+(-)^{|\Phi|}\underline{\Phi}{~\!}\underline{C}
\equiv\{\underline{C},\underline{\Phi}\}\\
8D\wedge\Phi&\leftrightarrow \underline{D}{~\!}\underline{\Phi}+(-)^{|\Phi|}\underline{\Phi}{~\!}\underline{D}
+\gamma^m\underline{\Phi}{~\!}\underline{D}_m+
(-)^{|\Phi|}\underline{D}_m\underline{\Phi}\gamma^m
\equiv\{\underline{D},\underline{\Phi}\}
~,}}
where $C$ is an arbitrary one-form and $D$ is an arbitrary three-form.
For concreteness let us now specialize to the case of IIA;
the proof for the case of IIB is essentially identical. Using (\ref{puregen}), (\ref{corr})
we see that
\eq{\label{int2}\d\Psi_2\leftrightarrow 16(
\gamma^m\nabla_m\eta_1\otimes\widetilde{\eta^c_2}+
\gamma^m\eta_1\otimes\nabla_m\widetilde{\eta^c_2}+
\nabla_m\eta_1\otimes\widetilde{\eta^c_2}\gamma^m+
\eta_1\otimes\nabla_m\widetilde{\eta^c_2}\gamma^m
)~,}
where without loss of generality we have parameterized the norm of the
supersymmetry parameters in terms of a real scalar $\alpha$: $\epsilon_1=\alpha\eta_1$, $\epsilon_2=\alpha\eta_2^c$ so that the
spinors $\eta_1$, $\eta_2$ are unimodular and of positive chirality. Moreover  the supersymmetry equations (\ref{kse}) imply
\eq{\spl{\label{int3}
\frac{1}{2}\underline{H}\eta_1&=
-\underline{\partial\phi}~\!\eta_1+
\frac{1}{16}e^\phi\gamma^m\underline{F}\gamma_m\eta_2^c\\
\frac{1}{2}\widetilde{\eta^c_2}\underline{H}&=
-\widetilde{\eta^c_2}\underline{\partial\phi}-
\frac{1}{16}e^\phi\widetilde{\eta_1}\gamma^m\underline{F}\gamma_m
\\
\nabla_m\eta_1&=
-\eta_1\partial_m\log\alpha-\frac{1}{4}\underline{H}_m\eta_1
+\frac{1}{16}e^\phi\underline{F}\gamma_m\eta_2^c\\
\nabla_m\widetilde{\eta^c_2}&=-\widetilde{\eta^c_2}\partial_m\log\alpha
-\frac{1}{4}\widetilde{\eta^c_2}\underline{H}_m
-\frac{1}{16}e^\phi\widetilde{\eta}_1\gamma_m\underline{F}
~.}}
Using (\ref{int3}) we can rewrite rewrite  (\ref{int2}) as
\eq{
\d\Psi_2\leftrightarrow
\frac{1}{2}\{\underline{\d\phi}-2\underline{\d\log\alpha},\underline{\Psi}_2\}
-\frac{1}{8}\{\underline{H},\underline{\Psi}_2\}
~,}
where we have taken into account the definitions of the brackets given in (\ref{int1}). Furthermore using the Clifford equivalences in (\ref{int1}) the equation above becomes
\eq{
\d_H\big(\alpha^2e^{-\phi}\Psi_2\big)\leftrightarrow 0
~,}
which is of course equivalent to the first line in (\ref{main}).

\section{Uplift to Lorentzian eleven-dimensional supergravity}\label{sec:uplift}

Massless  ($F_0=0$) {\it real} Euclidean IIA supergravity  in ten dimensions can be obtained from the reduction of eleven-dimensional Lorentzian supergravity on a timelike direction \cite{h}. More specifically: provided the reality conditions (\ref{realiia}), (\ref{realiiaspin}), (\ref{realiiat}) are imposed, the supersymmetric Euclidean IIA bosonic backgrounds of section \ref{sec:eiia} uplift to
$\mathcal{N}=1$ bosonic backgrounds of eleven-dimensional Lorentzian supergravity \cite{inta,Gutowski:2014ova} with eleven-dimensional metric
\eq{\label{mmetric}
\d s_{11}^2 = -e^{{\frac43}\phi}(C_1+\d t)^2+
e^{-{\frac23}\phi}\d s^2(\mcal_{10})
 ~,}
where
the Euclidean  ten dimensional metric $\d s^2(\mcal_{10})$ is in the string frame;
all fields are assumed time-independent and $C_1$ is the one-form potential for the
two-form fieldstrength of real Euclidean IIA supergravity:
\eq{F_2=i\d C_1~.}
Note that $C_1$ is real since, as follows from (\ref{realiia}), $F_2$ is imaginary.
The ten-dimensional Euclidean IIA supersymmetry parameters $\epsilon_1$, $\epsilon_2$ uplift to a single $Spin(1,10)$ eleven-dimensional supersymmetry parameter $\epsilon$ given by
\eq{\label{52}
\epsilon=\epsilon_1+\epsilon_2
~.}
The spinor $\epsilon$ is Majorana
as follows from (\ref{realiiaspin}) and the discussion around (\ref{real}).
The eleven-dimensional four-form $G_4$ is given by
\eq{\label{53}G_4=F_4+i(C_1 + \d t)\wedge H~,}
where $H$ is the NSNS three-form of real Euclidean IIA. From (\ref{realiia}) it follows that $H$ is imaginary and hence $G_4$ is real, as it should.


\subsection{Integrability}\label{sec:int}

It can immediately be seen that the Bianchi identity for the four-form $G_4$
\eq{\label{5a}\d G_4=0~,}
is equivalent to the five-form part of the RR Bianchi identity together with the Bianchi identity for the NSNS form
\eq{\d F_4+H\wedge F_2=0~; ~~~\d H=0~.}
Moreover the four-form equation of motion
\eq{\label{5b}\d\star_{11} G_4-\frac12 G_4\wedge G_4=0~,}
where the Hodge star above is taken with respect to the eleven-dimensional metric, is equivalent to the seven-form part of the RR Bianchi identity
together with the NSNS field equation of motion
\eq{\d F_6+H\wedge F_4=0~; ~~~\d\big( e^{-2\phi}\star_{10} H\big)
+\frac12 \big(\star_{10} F\wedge F\big)_8 =0~,}
where the Hodge star above is taken with respect to the Euclidean  ten-dimensional metric of $\mcal_{10}$ in the string frame. The eleven-dimensional supergravity
integrability theorem of \cite{inta,inta2,inta3} applied
to the supersymmetric solutions of this section implies that under certain
mild conditions\footnote{A sufficient condition is that the mixed time-space components of the Einstein equations are satisfied, however it can be seen that this will automatically be the case if the Killing vector
constructed as a bilinear of the Killing spinor is timelike \cite{inta}.} imposing
(\ref{5a}), (\ref{5b}) suffices to guarantee that all remaining
equations of motion (i.e. the eleven-dimensional Einstein equations) are
automatically satisfied.

\subsection{A conformal K\"{a}hler example}\label{sec:example}

A simple supersymmetric eleven-dimensional supergravity
solution can be obtained as the uplift of the IIA solution (\ref{iiasol}) with vanishing RR flux, $F=0$. The metric is given by
\eq{\d s_{11}^2=
-e^{{\frac43}\phi}\d t^2+
e^{-{\frac23}\phi}\d s^2(\mcal_{10})  ~,}
with $\mcal_{10}$ any manifold with torsion classes
\eq{\label{t}W_1=W_2=W_3=0~;~~~W_4=\frac12\partial^+\phi~;~~~W_5=\partial^+\phi~.}
As can be seen from (\ref{iiasol}) this is consistent with $f^{(1,0)}_4=0$ and the
vanishing of all ten-dimensional RR flux. Geometrically this condition means that $\mcal_{10}$ is a conformal K\"{a}hler manifold since all torsion classes except for $W_5$ can be made to vanish by a Weyl rescaling of the
vielbeine: $e_m^a\rightarrow e^{-\phi/4}e_m^a$.
In particular since $\mcal_{10}$ is complex we may introduce holomorphic coordinates and identify $\partial^+$, $\partial^-$ with the holomorphic, antiholomorphic differential respectively.

Moreover, as can be seen from (\ref{53}), (\ref{iiasol}), $H=\frac{i}{2}\d \phi\wedge J$ and the four-form flux is given by
\eq{G_4=-\frac12 \d t\wedge\d \phi\wedge J~.}
Taking into account that
\eq{\d J=\frac12\d\phi\wedge J~,}
which follows from (\ref{t}), (\ref{torj}), it can be see that the
Bianchi identity (\ref{5a}) is automatically satisfied.
Taking into account that the ten-dimensional Hodge dual of $\d\phi\wedge J$ is proportional to $J^3\wedge(\partial^+\phi-\partial^-\phi)$, it can also be seen that
the equation of motion (\ref{5b}) is equivalent to the condition
\eq{\partial^+\wedge\partial^-e^{-\phi/2}=0~.}
This is solved by
\eq{e^{-\phi/2}=f(z)+f(z)^*~,}
where $z$ denotes the holomorphic coordinates of $\cm_{10}$ and $f$ is an
arbitrary holomorphic function. If $\cm_{10}$ is compact $f$ must therefore be constant and the four-form flux vanishes\footnote{This is a
special case of the `vanishing theorem' shown in \cite{inta} which,  in the
notation of the present paper, states that
if $H=0$ and $\cm_{10}$ is compact, smooth and without a boundary then  $G_4$ vanishes.} while $\cm_{10}$ reduces to a
CY fivefold; this can of course be avoided by taking $\cm_{10}$ to
be non-compact, or by allowing higher-order corrections on the right-hand side
of (\ref{5b}).

\section*{Acknowledgment}

We would like to thank David Andriot and Jan Rosseel for useful discussions.

\appendix

\section{Spinor and gamma matrix conventions}\label{app1}

For a spinor $\psi$ in any dimension we define:
\eq{\widetilde{\psi}\equiv\psi^{Tr}C^{-1}~,}
where $C$ is the charge conjugation matrix. In Lorentzian signatures, we also define
\eq{\overline{\psi}\equiv\psi^{\dagger}\G_{0}~,}
where the Minkowski metric is mostly plus.
In all dimensions the Gamma matrices are taken to obey
\eq{
\G_M^{\dagger}=\G_0\G_M\G_0~.
}
Antisymmetric products of
Gamma matrices are defined by
\eq{
\G^{(n)}_{M_1\dots M_n}\equiv\G_{[M_1}\dots\G_{M_n]}~.
}

\subsection*{Ten Euclidean dimensions}

The charge conjugation matrix obeys:
\begin{align}
C^T = - C; \quad C^\dagger=  C^{-1} ; \quad C^* = - C^{-1}
\end{align}
The complex conjugate $\eta^c$ of a spinor $\eta$ is given by
\eq{\eta^c=C\eta^*~.}
The chirality operator is defined by:
\begin{align}
\g_{11} = i \g_1 ... \g_{10}~.
\end{align}
The irreducible spinor representations of $Spin(10)$ are given by  sixteen-dimensional  Weyl
spinors which are complex, in the sense that $\eta^c$ and
$\eta$ have opposite chiralities.
The Hodge dual of an antisymmetric product of $k$ gamma matrices is given by:
\begin{align}\label{eq:gamhodge}
\star \g_k = - i (-1)^{\frac{1}{2} k (k+1)} \g_{10-k} \g_{11}~.
\end{align}

\subsection*{Eleven Lorentzian dimensions}

Given a set gamma-matrices $\{\gamma_a\}$, $a=1,\dots 10$, generating the Clifford algebra in ten Euclidean dimensions the
eleven-dimensional Lorentzian gamma matrices are given by:
\eq{\label{gammauplift}
\Gamma_a=\left\{\begin{array}{rl}
i\gamma_{11}~;&~~~ a=0\\
\gamma_a~;&~~~ a=1,\dots 10\\
\end{array}\right.
~.}
In our conventions the charge conjugation matrix $C$ in eleven Lorentzian dimensions is the same as the one in ten Euclidean dimensions.

Consider a Dirac spinor $\epsilon$ of $Spin(1,10)$ (with 32 complex components).
Under $$Spin(1,10)\rightarrow Spin(10)~,$$
the spinor $\epsilon$ decomposes as $${\bf 32}\rightarrow{\bf 16_+}\oplus {\bf 16_-}~,$$ where ${\bf 16_{\pm}}$ are the positive-, negative-chirality Weyl spinors of $Spin(10)$ (with 16 complex components each). Explicitly we have:
\eq{\epsilon=\epsilon_1+\epsilon_2
~,}
where $\epsilon\sim{\bf 32}$ of $Spin(1,10)$, $\epsilon_1\sim{\bf 16_+}$ of $Spin(10)$ and $\epsilon_2\sim{\bf 16_-}$ of $Spin(10)$.
Imposing the Majorana condition on $\epsilon$,
\eq{\bar{\epsilon}=\widetilde{\epsilon}
~,}
is equivalent to
\eq{\label{real}\epsilon_2=-i\epsilon_1^c~,}
where we have taken (\ref{gammauplift}) into account.

\section{$SU(5)$ structures}

As discussed in the main text, a nowhere-vanishing  pure Weyl spinor
$\eta$ of unit norm in ten Euclidean dimensions defines an $SU(5)$
structure. In ten Euclidean dimensions not every Weyl spinor is pure:
the property of purity is equivalent to the condition
\eq{\widetilde{\eta}\gamma_m\eta=0~,}
for any gamma-matrix $\gamma_m$.

Let us  define a real two-form $J$ and a complex self-dual four-form $\Omega$ through the spinor bilinears
\eq{\label{su5}
i J_{mn} = \widetilde{\eta^c} \g_{mn} \eta ~;~~~
\O_{mnpqr} = \we \g_{mnpqr} \eta \;.
}
It can be shown by Fierzing that these forms obey:
\eq{\spl{\label{b4}
J\wedge\Omega&=0\\
\frac{i}{2^5}\Omega\wedge\Omega^*&=\frac{1}{5!}J^5 =\mathrm{vol}_{10}~,
}}
up to a choice of orientation,
and hence define an $SU(5)$ structure.
Raising one index of $J$ with the metric defines an almost complex structure:
\eq{
J_m{}^pJ_p{}^n=-\delta_m^n
~.}
Using the almost complex structure
we can define the projectors
\eq{
(\Pi^{\pm})_m{}^n\equiv\frac{1}{2}(\delta_{m}{}^{n}\mp i J_m{}^n)
~,}
with respect to which $\Omega$ is holomorphic
\eq{
(\Pi^{+})_m{}^i\Omega_{inpqr}=\Omega_{mnpqr}~; ~~~~~(\Pi^{-})_m{}^i\Omega_{inpqr}=0 ~.
}
The spinor bilinears are given by:
\eq{\spl{
\widetilde{\eta^c}\eta=1; &~~~~~\we \g_m \eta= 0 \\
\widetilde{\eta^c}\g_{mn}\eta=iJ_{mn}; &~~~~~\we\g_{mnp}\eta=0\\
 \widetilde{\eta^c}\g_{mnpq}\eta=-3J_{[mn}J_{pq]}; &~~~~~\we\g_{mnpqr}\eta=  \O_{mnpqr}\\
\widetilde{\eta^c}\g_{mnpqrs}\eta=-15iJ_{[mn}J_{pq}J_{rs]}; &~~~~~\we\g_{mnpqrst}\eta=0\\
\widetilde{\eta^c}\g_{mnpqrstu}\eta=105J_{[mn}J_{pq}J_{rs}J_{tu]}; &~~~~~\we\g_{mnpqrstuv}\eta=0 ~\\
\widetilde{\eta^c}\g_{mnpqrstuvw}\eta= 945 i J_{[mn}J_{pq}J_{rs}J_{tu}J_{vw]} ~,
}}
whereas the bilinears $\we\g_{(2p)}\eta$, ~$\widetilde{\eta^c}\g_{(2p-1)}\eta$, vanish. The following useful identities can be proved  by Fierzing
\eq{\spl{ \label{eq:omegas}
\frac{1}{5!\times 2^5}~&\Omega_{vwxyz}\Omega^{*vwxyz}=1\\
\frac{1}{4!\times 2^5}~&\Omega_{awxyz}\Omega^{*mwxyz}
=(\Pi^+)_{a}{}^{m}\\
\frac{1}{12\times 2^5}~&\Omega_{abxyz}\Omega^{*mnxyz}
=(\Pi^+)_{[a}{}^{m}(\Pi^+)_{b]}{}^{n}\\
\frac{1}{12\times 2^5}~&\Omega_{abcyz}\Omega^{*mnpyz}
=(\Pi^+)_{[a}{}^{m}(\Pi^+)_{b}{}^{n}(\Pi^+)_{c]}{}^{p}\\
\frac{1}{4!\times 2^5}~&\Omega_{abcdz}\Omega^{*mnpqz}
=(\Pi^+)_{[a}{}^{m}(\Pi^+)_{b}{}^{n}(\Pi^+)_{c}{}^{p}(\Pi^+)_{d]}{}^{q}\\
\frac{1}{5!\times 2^5}~&\Omega_{abcde}\Omega^{*mnpqr}
=(\Pi^+)_{[a}{}^{m}(\Pi^+)_{b}{}^{n}(\Pi^+)_{c}{}^{p}(\Pi^+)_{d}{}^{q}(\Pi^+)_{e]}{}^{r}\\
}}
Moreover:
\eq{\spl{\label{jvol}
\varepsilon_{mnpqrstuvw}J^{mn}J^{pq}J^{rs}J^{tu}J^{vw} &=3840\\
\varepsilon_{mnpqrstuvw}J^{pq}J^{rs}J^{tu}J^{vw}       &=384 J_{mn}\\
\varepsilon_{mnpqrstuvw}J^{rs}J^{tu}J^{vw}             &=144 J_{[mn}J_{pq]}\\
\varepsilon_{mnpqrstuvw}J^{tu}J^{vw}                   &=120  J_{[mn}J_{pq}J_{rs]}\\
\varepsilon_{mnpqrstuvw}J^{vw}                         &=210 J_{[mn}J_{pq}J_{rs}J_{tu]}\\
\varepsilon_{mnpqrstuvw}                               &=945 J_{[mn}J_{pq}J_{rs}J_{tu}J_{vw]}
~.
}}
The last line of the above equation together with the last line of \eqref{eq:omegas}
imply
\eq{\label{omvol}
\Omega_{[a_1\dots a_5}\Omega^*_{a_6\dots a_{10}]}=- \frac{8i}{63}\varepsilon_{a_1\dots a_{10}}~;~~~
\Omega_{a_1\dots a_5}=-  \frac{i}{5!}\varepsilon_{a_1\dots a_{10}}
\Omega^{a_6\dots a_{10}}
~.
}
%
%
%
Finally, the following relations are useful in the analysis of the Killing spinor equations.
\eq{\spl{
\g_m\eta      &= (\Pi^+)_{m}{}^{n}\g_n\eta\\
\g_{mn}\eta   &= iJ_{mn}\eta + (\Pi^+)_{[m}^{\phantom{[m}p}(\Pi^+)_{n]}^{\phantom{[m}q} \g_{pq}   \eta  \\
\g_{mnp}\eta  &= 3iJ_{[mn}\g_{p]}\eta  +\frac{1}{8}\Omega_{mnpqr}\g^{qr}\eta^c\\
\g_{mnpq}\eta &= -3J_{[mn}J_{pq]}\eta + 6i J_{[mn}(\Pi^+)_{p}^{\phantom{[m}r}(\Pi^+)_{q]}^{\phantom{[m}s} \g_{rs}   \eta
                 - \frac{1}{2} \O_{mnpqr} \g^r \eta^c \\
\g_{mnpqr}\eta &=  - \O_{mnpqr} \eta^c + \frac{5i}{4} J_{[mn}\O_{pqrst}\g^{st} \eta^c - 15 J_{[mn}J_{pq}\g_{r]} \eta
~.
}}

\subsection{Torsion classes}

The intrinsic torsion $\tau$ (see \cite{intrinsic} for a review) is an element of  $\Lambda^1(T^*\mathcal{M}) \otimes su(5)^\perp$, with  $su(5)^\perp \oplus su(5) = so(10)$. Since
\eq{\spl{
\tau &\in(\bf{5}\oplus\bf{\bar{5}})\otimes(\bf{1}\oplus \bf{10}\oplus\bf{\bar{10}})\nn\\
&\sim(\bf{10}\oplus\bf{\bar{10}})\oplus(\bf{40}\oplus\bf{\bar{40}})
\oplus(\bf{45}\oplus\bf{\bar{45}})
\oplus(\bf{5}\oplus\bf{\bar{5}})\oplus(\bf{5}\oplus\bf{\bar{5}})
~,}}
we can decompose the intrinsic torsion in terms of the torsion classes $W_1,\dots, W_5$, which are irreducible representations of $su(5)$. These torsion classes are the obstructions to the closure of the forms $J$, $\Omega$. Explicitly we will choose the following parameterization:\footnote{We define the contraction
between a $p$-form $\varphi$ and a $q$-form $\chi$, $p\leq q$, by
\eq{
\varphi\lrcorner\chi=\frac{1}{p!(q-p)!}\varphi^{m_1\dots m_p}
\chi_{m_1\dots m_p n_1\dots n_{q-p}}\d x^{n_1}\wedge\dots\wedge\d x^{n_{q-p}}
\nn~.}
Once the normalization of the $W_1$ term on the right-hand side of the first equation in (\ref{torj}) is fixed, the $W_1$ term on the right-hand side of the second equation can be determined as follows: Starting from $\d(J\wedge\Omega)=0$ we substitute for $\d J$, $\d\Omega$ using (\ref{torj}), taking (\ref{jvol}), (\ref{omvol}) into account and noting that
$W_2\wedge J\wedge J=0$ since $W_2$ is primitive.}
\eq{\spl{\label{torj}
\d J&=W^*_1\lrcorner\Omega+W_3+W_4\wedge J+\mathrm{c.c.}\\
\d\Omega&= - \frac{16i}{3} W_1\wedge J\wedge J+W_2\wedge J+W^*_{5}\wedge\Omega
~,}}
where $W_1 \sim \bf{10}$ is a complex (2,0)-form, $W_2 \sim \bf{40}$ is a complex primitive (3,1)-form, $W_3 \sim \bf{45}$ is a complex primitive (2,1)-form  and $W_4$, $W_5\sim\bf{5}$ are complex (1,0)-forms.

Since the $SU(5)$ structure $(J,\O)$ can be expressed in terms of the spinor bilinears (\ref{su5}), the torsion classes can also be interpreted as obstructions to $\eta$ being covariantly constant. Explicitly we have:
\eq{\label{tornabl}\spl{
\nabla_m \eta =& \left(W_4 - \frac{1}{2}W_5 - \mathrm{c.c.}\right) \eta \\
&+ \left( - \frac{i}{48} \O_{mnpqr}^*W_1^{qr} + \frac{1}{4}
\Pi^+_{m[n}W_{4p]}^* - \frac{i}{8}  W_{3mnp}^* + \frac{i}{4!2^5}  W_{2mxyz} \O_{np}^{*\phantom{np}xyz} \right) \g^{np} \eta ~.}}
This can be seen as follows. In $D=10$, the Weyl spinor $\eta$ has 16 complex degrees of freedom. Due to transitivity of the Clifford algebra, we can thus express any Weyl spinor $\xi$ of positive chirality as
\eq{
\xi = \chi \eta + \chi_n \g^n \eta^c + \chi_{np} \g^{np} \eta
~,}
where $\chi\sim{\bf 1}$ is a complex scalar, $\chi_n\sim{\bf 5}$ is a (1,0)-form and $\chi_{np}\sim{\bf \bar{10}}$ is a (0,2)-form, and similarly for negative-chirality spinors. As a consistency check we note that the arbitrary positive-chirality spinor $\xi$ is parametrized by sixteen complex degrees of freedom: one complex d.o.f. from the complex scalars $\chi$, five complex d.o.f.s from the complex (1,0)-form  $\chi_{n}$, plus ten complex d.o.f.s from the complex (0,2)-form  $\chi_{np}$.
In the same way we can express the covariant derivative of $\eta$ as
follows
\eq{\label{b10}
\nabla_m \eta = \varphi_m \eta + \varphi_{m,n} \g^n \eta^c + \varphi_{m,np} \g^{np} \eta \;,
}
for some complex coefficients $\varphi_m\sim (\bf{5} \oplus \bf{\bar{5}} )$, $\varphi_{m,n}\sim(\bf{5} \oplus \bf{\bar{5}} )\otimes\bf{5}$,
$\varphi_{m,np}\sim (\bf{5} \oplus \bf{\bar{5}} )\otimes\bf{\bar{10}}$.
Moreover, the purity of $\eta$ implies $\widetilde{\eta}\gamma_m\nabla_n\eta=0$ and thus $\varphi_{m,n}=0$. Similarly the constancy of the norm of $\eta$
implies that $\varphi_m$ is imaginary: $\varphi^{(1,0)*}_m = - \varphi^{(1,0)}_m$.
We can further decompose:
\eq{
(\bf{5} \oplus \bf{\bar{5}} )\otimes \bar{10}= \bf{\bar{5}} \oplus  \bf{10} \oplus \bf{40} \oplus \bf{45}~,
}
which explicitly amounts to parameterizing
\eq{\label{b11}
\varphi_{m,pq} = \O_{mpq}^{*\phantom{mpq}ab}E_{ab}^{(2,0)} + \Pi^+_{m[p}F^{(0,1)}_{q]} + G_{mpq}^{(1,2)} + H^{(3,1)}_{mabc}\O^{*abc}_{\phantom{*abc}pq} \;,
}
where now all coefficients on the right-hand side above are in irreducible
$su(5)$ modules. Taking the above into account we can now multiply (\ref{b10}) on the left with $\widetilde{\eta^c}\g_{ij}$ and $\we\g_{ijklr}$, and antisymmetrize in all free indices  in order to form $\d J$ and $\d \Omega$ respectively as spinor bilinears. Comparing with (\ref{torj}) then leads to \eqref{tornabl}.

\subsection{Tensor decomposition}\label{sec:tensor}

Under an $so(10)\rightarrow su(5)$ decomposition the one-form, the two-form, the three-form, the four-form, and the self-dual five-form of $so(10)$ decompose respectively as:
\eq{\spl{
\bf{10}   &\rightarrow \bf{ 5\oplus \bar{5} }\\
\bf{45}   &\rightarrow \bf{1 \oplus 10 \oplus \bar{10} \oplus  24}\\
\bf{120}  &\rightarrow \bf{5\oplus \bar{5} \oplus 10 \oplus \bar{10} \oplus 45\oplus \bar{45} }\\
\bf{210}  &\rightarrow \bf{ 1\oplus 5\oplus \bar{5} \oplus 10 \oplus \bar{10} \oplus 24 \oplus  40 \oplus \bar{40} \oplus 75 }\\
\bf{126^+} &\rightarrow 1 \oplus \bf{5}\oplus \bar{10} \oplus \bar{15}  \oplus\bar{45}\oplus   50
~.
}}
Note that  the $\bf{126^+}$ is an imaginary self-dual five-form as defined in (\ref{sd}): $\star F_5 = i F_5$.
 \\
Explicitly we decompose the RR forms as follows:
\eq{\spl{
F_m        &= f_{1|m}^{(1,0)} + f_{1|m}^{(0,1)}                \\
F_{mn}     &= f_2 J_{mn} + f^{(1,1)} + f_{2|mn}^{(2,0)} + f_{2|mn}^{(0,2)}   \\
F_{mnp}    &= 3 f_{3|[m}^{(1,0)} J_{np]} + f_{3|mnp}^{(2,1)} + \frac{1}{2} {f}^{(0,2)}_{3|qr} \O_{mnp}^{\phantom{mnp}qr} +3 f_{3|[m}^{(0,1)} J_{np]} + f_{3|mnp}^{(1,2)} + \frac{1}{2} {f}^{(2,0)}_{3|qr} \O_{mnp}^{*\phantom{mnp}qr} \\
F_{mnpq}   &= 6 f_4 J_{[mn}J_{pq]} + 6 f_{4|[mn}^{(1,1)}J_{pq]} + f_{4|mnpq}^{(2,2)} \\
            &+ 6 ( f_{4|[mn}^{(2,0)} + f_{4|[mn}^{(0,2)})  J_{pq]} + f_{4|mnpq}^{(3,1)} + f_{4|mnpq}^{(1,3)}
           + {f}^{(0,1)}_{4|r} \O_{mnpq}^{\phantom{mnpq}r} + {f}^{(1,0)}_{4|r} \O_{mnpq}^{*\phantom{mnpq}r} \\
F^+_{mnpqr}&= 30 f_{5|[m}^{(1,0)} J_{np} J_{qr]} + 5 {f}^{(0,2)}_{5|xy} \O_{[mnp}^{\phantom{mnp}xy} J_{qr]}
                 + 10 f^{(1,2)}_{5|[mnp}J_{qr}] + {f}_5 \O_{mnpqr} + f^{(1,4)}_{5|mnpqr} + f^{(3,2)}_{5|mnpqr}  \;,
}}
where in terms of irreducible $su(5)$ representations we have:
\eq{\spl{
f                 &\sim \bf{1}    \hskip .32cm;  \\
f_{mn}^{(2,0)}    &\sim \bf{10}   \;;   \\
f_{mnp}^{(2,1)}   &\sim \bf{45}   \;;    \\
f_{mnpq}^{(3,1)}  &\sim \bf{40}   \;;     \\
f_{mnpqr}^{(4,1)} &\sim \bf{15}   \;;
}
\quad
\spl{
f_m^{(1,0)}       &\sim \bf{5} \\
f_{mn}^{(1,1)}    &\sim \bf{24} \\
& \\
f_{mnpq}^{(2,2)}  &\sim \bf{75}  \\
f_{mnpqr}^{(3,2)} &\sim \bf{50}   \;.
}
}
Unless otherwise stated all forms are a priori complex. Demanding reality of the $so(10)$ representations would be equivalent to imposing $(f^{(p,q)})^* = f^{(q,p)}$.

Similarly for the NSNS three form we expand:
\eq{H_{mnp}    = 3 h_{3|[m}^{(1,0)} J_{np]} + h_{3|mnp}^{(2,1)} + \frac{1}{2} {h}^{(0,2)}_{3|qr} \O_{mnp}^{\phantom{mnp}qr} +3 h_{3|[m}^{(0,1)} J_{np]} + h_{3|mnp}^{(1,2)} + \frac{1}{2} {h}^{(2,0)}_{3|qr} \O_{mnp}^{*\phantom{mnp}qr} ~.}
Using Hodge duality we can similarly express all $p$-forms with $p > 5$ in terms of the expansions above; the following expressions are needed
in section \ref{iiagen}
\eq{\spl{
F_6 &= -\frac{1}{3} i f_4J \wedge J \wedge J + \frac12 i (f_4^{(1,1)}- f_4^{(2,0)}- f_4^{(0,2)} )\wedge J \wedge J \\
&+ i (f_4^{(3,1)}+f_4^{(1,3)} -f_4^{(2,2)} ) \wedge J -  \tilde{f}_4^{(0,1)} \wedge \O  +  \tilde{f}_4^{(1,0)} \wedge \O^* \\
F_8 &= i \frac{1}{4!}f_2 J^4 + i\frac{1}{3!}(f_2^{(2,0)}+ f_2^{(0,2)}- f_2^{(1,1)})J^3 \\
F_{10} &= -i \frac{1}{5!}f_0 J^5\;.
}}

\subsection{Useful formul\ae{}}

In this appendix we list some intermediate formul\ae{} which
are useful in the analysis of the Killing spinor equations.

\subsection*{Formul\ae{} needed for the dilatino equations in IIA}

\eq{\spl{
\underline{F_2} \eta &= 5 i f_2 \eta + \frac{1}{2}f_{2|mn}^{(0,2)} \g^{mn} \eta \\
\underline{H} \eta &= 4 i h_m^{(0,1)} \g^m \eta + 4 {h}_{mn}^{(2,0)} \g^{mn} \eta^c \\
\underline{F_4} \eta &= - 20 f_4 \eta +  \frac{3i}{2} f_{4|mn}^{(0,2)} \g^{mn} \eta  - 16 \tilde{f}_{4|m}^{(1,0)} \g^m \eta^c
~.}}

\subsection*{Formul\ae{} needed for the gravitino equations in IIA}
\eq{\spl{
\underline{F_2}\g_m \eta &= 3 i f_2 \g_m \eta - 2 f_{2|mn}^{(1,1)} \g^n \eta + \frac{1}{16} f_{2|qr}^{(0,2)} \O_{mnp}^{\phantom{mnp}qr} \g^{np} \eta^c\\
\underline{H}_m \eta &= 4 i \left( h_m^{(0,1)} +h_m^{(1,0)} \right)\eta + i (\Pi^+)_{mn} h_p^{(0,1)}
                        \g^{np}\eta + \frac{1}{2} h_{mnp}^{(1,2)}\g^{np} \eta
                        + \frac{1}{4} {h}^{(2,0)}_{qr} \O_{mnp}^{*\phantom{mnp}qr}\g^{np} \eta
\\
\underline{F_4} \g_m \eta &= - 4 f_4 \g_m \eta - 6 i  f_{4|mn}^{(1,1)} \g^n \eta - 32 {f}_{4|m}^{(1,0)} \eta^c
                             - \frac{1}{24}  f_{4|mqrs}^{(1,3)} \O_{np}^{\phantom{np}qrs} \g^{np} \eta^c
                                                  + \frac{i}{16}  f^{(0,2)}_{4|qr} \O_{mnp}^{\phantom{mnp}qr} \g^{np} \eta^c
~.}}

\subsection*{Formul\ae{} needed for the dilatino equations in IIB}
\eq{\spl{
\underline{F_1}\eta &= f_{1|m}^{(0,1)} \g^m \eta \\
\underline{F_3}\eta &= 4i f_{3|m}^{(0,1)} \g^m \eta + 4 {f}_{3|mn}^{(2,0)} \g^{mn} \eta^c
~.}}

\subsection*{Formul\ae{} needed for the gravitino equations in IIB}
\eq{\spl{
\underline{F_1} \g_m \eta &= 2f_{1|m}^{(1,0)}\eta - \Pi^+_{mn} f_{1|p}^{(0,1)} \g^{np} \eta              \\
\underline{F_3} \g_m \eta &= 8i f_{3|m}^{(1,0)}\eta    -2i \Pi^+_{mn}f_{3|p}^{(0,1)} \g^{np} \eta
                           - 16  {f}_{3|mn}^{(2,0)} \g^{n} \eta^c + f_{3|mnp}^{(1,2)} \g^{np} \eta  \\
\underline{F_5} \g_m \eta &= - 24 f_{5|m}^{(1,0)} \eta
                             + 2i  f_{5|mnp}^{(1,2)} \g^{np} \eta
                             - \frac{1}{4!} f_{5|mnpqr}^{(1,4)} \O^{npqrs} \g_s \eta^c
~.}}

\subsection*{Formu\ae{} needed for the generalized geometric computations in IIA}
\eq{\spl{\label{for}
(h^{(0,2)}\lrcorner \Omega)\wedge\Omega^*&=-\frac{16i}{3}h^{(0,2)}\wedge J^3\\
(\tilde{f}_4^{(1,0)} \lrcorner \O^* )\wedge J &=  i \tilde{f}_4^{(0,1)} \wedge \O^*
~.}}

\section{Mukai pairings}\label{sec:mukai}

The Mukai pairing is an inner product on the space of polyforms defined by
\eq{
\langle\Phi_1,\Phi_2\rangle=\left.\Phi_1\wedge\sigma(\Phi_2)\right|_{10}
~,}
where on the right-hand side above we have projected on the ten-form part of the polyform.
Using the polyform-bispinor correspondence this can also be written as
\eq{
\frac{1}{2^5}\mathrm{tr}(\underline{\widetilde{\Phi}_1}
\gamma_{11}\underline{\Phi_2})
\mathrm{vol}_{10}=
i(-)^{|\Phi_1|}
\langle\Phi_1,\Phi_2\rangle
~,}
where $(-)^{|\Phi|}$ is a positive, negative sign for $\Phi$ an even, odd polyform respectively; $\underline{\Phi}$ is the image of $\Phi$ under the Clifford map:
\eq{\label{cliff}\Phi\leftrightarrow\underline{\Phi}=
\sum_{p=0}^{10}\frac{1}{p!}\Phi_{m_1\dots m_p}\gamma^{m_1\dots m_p}
~,}
and we have defined:
\eq{\widetilde{\underline{\Phi}}\equiv
C\underline{\Phi}^{\mathrm{Tr}}C^{-1}
~.}
In ten Euclidean dimensions the Mukai pairing satisfies the following identities:
\eq{\label{c5}
\langle\star\sigma(\Phi_1),\Phi_2\rangle=-\langle\Phi_1,\star\sigma(\Phi_2)\rangle
=\langle\star\sigma(\Phi_2),\Phi_1\rangle
~,}
and
\eq{\label{c6}
\underline{\star\sigma(\Phi)}=i\gamma_{11}\underline{\Phi}
=i(-)^{|\Phi|}\underline{\Phi}\gamma_{11}
~.}
Applying (\ref{c6}) to the explicit Hodge diamond basis of section \ref{iiagen}
implies
\eq{\star\sigma(u^{(k,l)})=i(-)^{\frac{5-k-l}{2}}u^{(k,l)}
~.}
Using the above together with (\ref{c5}), (\ref{sd}) leads to the
following `selection rules':
\eq{
\big(1+ (-)^{\frac{5-l-k}{2}}\big)\langle F,u^{(k,l)}\rangle
=0~.}
This implies for example that $\langle F,u^{(k,1)}\rangle$ vanishes automatically for $k=0,\pm 4$, but not for $k=\pm 2$. For the computation of section \ref{sec:eiia} the following explicit Mukai pairings are useful:
\eq{\spl{\label{usefulmukai1}
\langle f\wedge J^2,u_{ab,c}^{(-2,1)}\rangle
&=-if^{ij}\Omega^*_{ijabc}\mathrm{vol}_{10}\\
\langle f^*\wedge J^2,u_{ab,c}^{(2,1)}\rangle
&=-if^{*ij}\Omega_{ijabc}\mathrm{vol}_{10}\\
\langle f\wedge J^3,u_{ab,c}^{(-2,1)}\rangle
&=-3f^{ij}\Omega^*_{ijabc}\mathrm{vol}_{10}\\
\langle f^*\wedge J^3,u_{ab,c}^{(2,1)}\rangle
&=-3f^{*ij}\Omega_{ijabc}\mathrm{vol}_{10}
~,}}
for any (2,0)-form $f$ transforming in the ${\bf 10}$ of $SU(5)$,
\eq{\spl{\label{usefulmukai2}
\langle f\wedge J,u_{ab,c}^{(-2,1)}\rangle
&=\frac13 f_{cijk}\Omega^*_{ab}{}^{ijk}\mathrm{vol}_{10}\\
\langle f^*\wedge J,u_{ab,c}^{(2,1)}\rangle
&=-\frac13 f^*_{cijk}\Omega_{ab}{}^{ijk}\mathrm{vol}_{10}
~,}}
for any traceless (3,1)-form $f$ in the ${\bf 40}$ of $SU(5)$,
\eq{\spl{\label{usefulmukai3}
\langle f,u_{a_1\dots a_4,b}^{(0,3)}\rangle
&=if \Omega_{a_1\dots a_4b}\mathrm{vol}_{10}\\
\langle fJ,u_{a_1\dots a_4,b}^{(0,3)}\rangle
&=3f \Omega_{a_1\dots a_4b}\mathrm{vol}_{10}\\
\langle f J^2,u_{a_1\dots a_4,b}^{(0,3)}\rangle
&=-4if \Omega_{a_1\dots a_4b}\mathrm{vol}_{10}
~,}}
for any scalar $f$ in the ${\bf 1}$ of $SU(5)$ and
\eq{\spl{\label{usefulmukai4}
\langle f,u_{a_1\dots a_4,b}^{(0,3)}\rangle
&=2if_b{}^c \Omega_{a_1\dots a_4c}\mathrm{vol}_{10}\\
\langle f\wedge J,u_{a_1\dots a_4,b}^{(0,3)}\rangle
&=6f_b{}^c \Omega_{a_1\dots a_4c}\mathrm{vol}_{10}
~,}}
for any traceless (1,1)-form $f$ in the ${\bf 24}$ of $SU(5)$.


\end{document}